\documentclass[]{raa}            

\usepackage{graphicx,times}  
\usepackage{natbib}
\usepackage{amssymb,amsmath}
\usepackage{amsmath}
\usepackage{multirow}
\usepackage{tabularx}
\usepackage{booktabs}
\usepackage[flushleft]{threeparttable}
\usepackage{lipsum}
\usepackage{enumitem}
\usepackage{hyperref}
\usepackage{siunitx}
\usepackage{rotating}
\usepackage{adjustbox} 
\bibpunct{(}{)}{;}{a}{}{,}

\usepackage{tikz,xcolor,hyperref}

\definecolor{lime}{HTML}{A6CE39}
\DeclareRobustCommand{\orcidicon}{%
	\begin{tikzpicture}
		\draw[lime, fill=lime] (0,0) 
		circle [radius=0.16] 
		node[white] {{\fontfamily{qag}\selectfont \tiny ID}};
		\draw[white, fill=white] (-0.0625,0.095) 
		circle [radius=0.007];
	\end{tikzpicture}
	\hspace{-2mm}
}

\foreach \x in {A, ..., Z}{%
	\expandafter\xdef\csname orcid\x\endcsname{\noexpand\href{https://orcid.org/\csname orcidauthor\x\endcsname}{\noexpand\orcidicon}}
}

\begin{document}

  \title{New cases of super-flares on slowly rotating solar-type stars and large amplitude super-flares in G- and M-type main-sequence stars.
}

   \volnopage{Vol.0 (20xx) No.0, 000--000}      
   \setcounter{page}{1}          

   \author{A. k. Althukair
      \inst{1,2}\orcidA{}
   \and D. Tsiklauri
      \inst{1}\orcidB{}
      }

   \institute{Department of Physics and Astronomy, School of Physical and Chemical Sciences, Queen Mary University of London,
   	Mile End Road, London, E1 4NS,
   	UK; {\it a.k.althukair@qmul.ac.uk}, {\it d.tsiklauri@qmul.ac.uk}\\
        \and
             Physics Department, College of Sciences, Princess Nourah Bint Abdulrahman University, Riyadh, PO Box 84428, Saudi Arabia\\
\vs\no
   {\small Received 20xx month day; accepted 20xx month day}}

\abstract{In our previous work, we searched for super-flares on different types of stars while focusing on G-type dwarfs using entire Kepler data to study statistical properties of the occurrence rate of super-flares. Using these new data, as a by-product, we found fourteen cases of super-flare detection on thirteen slowly rotating Sun-like stars with rotation periods of 24.5 to 44 days. This result supports earlier conclusion by others that the Sun may possibly have a surprise super-flare. Moreover, we found twelve and seven new cases of detection of exceptionally large amplitude super-flares on six and four main-sequence stars of G- and M-type, respectively. No large-amplitude flares were detected in A, F, or K main-sequence stars. Here we present preliminary analysis of these cases. The super-flare detection, i.e. an estimation of flare energy, is based on a more accurate method compared to previous studies. We fit an exponential decay function to flare light curves and study the relation between  e-folding decay time, $\tau$, vs. flare amplitude and flare energy. We find that for slowly rotating Sun-like stars, large values of $\tau$ correspond to small flare energies and small values of $\tau$ correspond to high flare energies considered. Similarly, $\tau$ is large for small flare amplitudes and $\tau$ is small for large amplitudes considered. However, there is no clear relation between these parameters for large amplitude super-flares in the main sequence  G- and M-type stars, as we could not establish clear functional dependence between the parameters via standard fitting algorithms.
}
   \authorrunning{Althukair \& Tsiklauri }            
   \titlerunning{Super-flares on Sun-like stars}  

   \maketitle

%
%
\section{Introduction}
It is believed the Solar and stellar flares are powered by
a physical process called magnetic reconnection, in which
connectivity of magnetic field lines in the atmospheres of
stars changes rapidly \citep{Masuda_1994,Shibata_1995}.  
This is accompanied by acceleration of
plasma particles and release of heat. The source of this
kinetic and thermal energy is the energy stored in the magnetic field.
Thus, the magnetic dynamo process which is one of possible
means to generate magnetic field via bulk plasma flows is
of great importance for understanding of what can power and
therefore be source for a flare or a super-flare.
The energies of observed stellar
flares lie in the wide range from $10^{28}$ to $10^{37}$ erg,
while the highest energy of any observed solar flare is
approximately few times $10^{32}$ erg. Thus, generally
agreed terminology is that a super-flare should have an 
energy in excess of $10^{34}$ erg.
A plausible dynamo model capable to explain 
the generation of magnetic
energy sufficient to support super-flares has been recently
suggested in \citet{ko16} and then 
further investigated in \citet{katsova18}.
In this scenario, rather than producing stellar cycles similar to the solar 11 year cycle, the dynamos in superflaring stars excite some 
quasi-stationary magnetic configuration with a
much higher magnetic energy. 
Further, \citet{kmn18} used a flux-transport model for the 
solar dynamo with fluctuations of the Babcock-Leighton type $\alpha$-effect 
to generate statistics of magnetic cycles. As a result, they concluded that 
the statistics of the computed energies of the cycles suggest 
that super-flares with energies in excess of $10^{34}$ erg are not possible on the Sun. 

Historical records suggest that no super-flares have occurred on the Sun in the last two millennia.
In the past there were notable examples detection of super-flares on Sun-like stars.
There are two references which support such clam \citet{Schaefer00} and \citet{Nogami14}. 

As claimed by \citet{Schaefer00} they identified nine cases of super-flares involving $10^{33}$ to $10^{38}$ ergs on main sequence Sun-like stars. 
Sun-like means that stars are on or near the main-sequence, 
have spectral class from F8 to G8, are single (or have a very distant binary companion).
The super-flare energy estimation by \citet{Schaefer00} was based on photometric methods.

\citet{Nogami14} reported  the results of high dispersion spectroscopy of two 'super-flare stars', KIC 9766237, and KIC 9944137 using Subaru/HDS telescope. 
These two stars are G-type main sequence stars, and have rotation 
periods of 21.8 days, and 25.3 days, respectively. 
Their spectroscopic results confirmed that these stars have stellar parameters similar to those of the Sun in terms of the effective temperature, surface gravity, and metallicity. 
By using the absorption line of Ca II 8542, 
the average strength of the magnetic field on the surface of these stars was estimated to be 1-20 G.  
The super-flare energy estimation by \citet{Nogami14} was based semi-empirical method based
on magnetic energy density times volume of flare.
These results claim that the spectroscopic properties of these super-flare stars are very close to those of the Sun, and support the hypothesis that the Sun may have a super-flare. 
What causes super-flares is an open issue and 
many theories exist to explain their origin.
\citet{Karak2020} have shown that sun-like slowly rotating stars, having anti-solar differential rotation, i.e.
when equatorial regions of the star rotate slower than the polar regions, can produce a very strong magnetic field and that could be a possible explanation for the superflare. 
It is the anti-solar differential rotation that can produce strong fields in slowly rotating stars.  
A study conducted by \citet{Karak2020} focus on mean-field kinematic dynamo modeling to investigate the behaviour of large-scale magnetic fields in different stars with varying rotation periods. They specifically consider two cases: stars with rotation periods larger than 30 days, which exhibit antisolar differential rotation (DR), and stars with rotation periods shorter than 30 days, which exhibit solar-like DR. 
The study supports the possible existence of antisolar differential rotation in slowly rotating stars and suggests that these stars may exhibit unusually enhanced magnetic fields and potentially produce cycles that are prone to the occurrence of super-flares.
In general the transition from solar to anti-solar differential rotation happens somewhere around the Rossby number of unity. For the Sun, it is 
obviously solar-like, but when the rotation rate decreases, one expects
to have an anti-solar differential rotation.  This robust transition has been seen in many numerical simulations.
\citet{Karak2015}, for example, using global MHD convection simulations,  consistently find anti-solar differential rotation 
when the star rotates slowly.

Statistical study of super-flares on different stellar types has been an active area of research \citep{maehara2012,Shibayama_2013,Wu2015,He2015,He2018,yang2017,Van-Doorsselaere2017,Lu2019,Yang2019,gunther2020,Tu2020,Gao2022}.
\citet{Shibayama_2013} studied statistics of stellar super-flares.
These authors discovered that for 
Sun-like stars (with surface temperature 5600-6000 K and 
slowly rotating with periods longer than 10 days), the occurrence rate of super-flares with an energy of 
$10^{34}-10^{35}$ erg is once in 800-5000 yr.
\citet{Shibayama_2013} confirmed the previous
results of \citet{maehara2012} in that the occurrence rate ($dN/dE$) of super-flares versus flare energy $E$ shows a power-law distribution
with $dN/dE \propto E^{-\alpha}$, where $\alpha \sim 2$. 
Such occurrence rate distribution versus flare 
energy is roughly similar to that for solar flares. \citet{Tu2020} identified and verified 1216 super-flares on 400 solar-type stars by analyzing 2-minute cadence data from 25,734 stars observed during the first year of the TESS mission. The results indicate a higher frequency distribution of super-flares compared to the findings from the Kepler mission. This difference may be due to a significant portion of the TESS solar-type stars in the dataset are rapidly rotating stars. The power-law index $\gamma$ of the super-flare frequency distribution was determined to be $\gamma = 2.16 \pm 0.10$, which is consistent with the results obtained from the Kepler mission. The study highlights an extraordinary star, TIC43472154, which exhibits approximately 200 super-flares per year. \citet{Tu2020} analyzed the correlation between the energy and duration of super-flares, represented as $T_{duration}\propto E^{\beta}$. They derived a power-law index $\beta = 0.42  \pm 0.01$ for this correlation, which is slightly larger than the value of $\beta = 1/3$ predicted by magnetic reconnection theory. 
Similar conclusion has been reached earlier by
\citet{Maehara2015}, who found that the duration of 
superflares, $\tau$, scales as the flare energy, 
$E$, according to $\tau \propto E^{0.39\pm 0.03}$.
\citet{Yang2023} analyzed TESS light curves from the first 30 sectors of TESS data with a two-minute exposure time. They identified a total of 60810 flare events occurring on 13478 stars and performed a comprehensive statistical analysis focusing on the characteristics of flare events, including their amplitude, duration, and energy.
We believe that method for flare energy estimation used in \citep{Shibayama_2013,yang2017}
is more accurate than one used by \citet{Schaefer00} and \citet{Nogami14}.
We therefore base of flare energy estimate on method of \citet{Shibayama_2013} and \citet{paper1} referred thereafter as Paper I.

In Paper I we searched for super-flares on 
different spectral class stars, while focusing on 
G-type dwarfs (solar-type stars)
using Kepler data using quarters $0-17$ with the purpose of study the 
statistical properties of the occurrence rate of super-flares. 
\citet{Shibayama_2013} studied statistics of stellar super-flares 
based on Kepler data in quarters $0-6$ ($Q0-Q6$). In Paper I we investigated
how the results are modified by adding more quarters, 
i.e. what is $\alpha$ power-law for data quarters $0-17$ and $7-17$.
Here using the more extended Kepler data, we also found 14 cases of detection of Super-flares on 13 Slowly Rotating Sun-like starts in each of KIC 3124010, KIC 3968932, KIC 7459381, KIC 7459381, KIC 7821531, KIC 9142489, KIC 9528212, KIC 9963105, KIC 10275962, KIC 11086906, KIC 11199277, KIC 11350663 and KIC 11971032. Thus the main purpose
of the present study is to present analysis of these new 14 cases. The main novelty here is that
the detection is based on a \citet{Shibayama_2013} method, which is more accurate,
as compared to \citet{Schaefer00} and \citet{Nogami14} for the
flare energy estimation. 
Our results support earlier conclusion by  
others (\citet{Schaefer00} and \citet{Nogami14}) that  
the Sun may have a surprise super-flare. We stress that Paper I
has conducted a more comprehensive study of determination of
stellar rotation periods based on a robust method such as
used by \citet{McQuillan2014} in
comparison to \citet{Shibayama_2013}.
We believe that more accurate period determination used in
Paper I has led to the current new results, presented in this paper. In addition to the 14 cases of of Super-flares on Slowly Rotating Sun-like starts, we detected 12 and 7 super-flares with a large amplitude on five G-type and four M-type main sequence stars, respectively.

Solar flares emit energy at all wavelengths, but their spectral distribution is still unknown. When white-light continuum emission is observed, the flares are referred to as "white-light flares" (WLF). \citet{Kretzschmar2011} identified and examined visible light emitted by solar flares and found that the white light is present on average during all flares and must be regarded as a continuum emission. \citet{Kretzschmar2011} also demonstrated that this emission is consistent with a black body spectrum with temperature of 9000 K and that the energy of the continuum contains roughly 70\% of the total energy emitted by the flares. WLFs are among the most intense solar flares, and it has been demonstrated that an optical continuum appears anytime the flare's EUV or soft X-ray luminosity reaches a reasonably large threshold \citep{McIntosh.Donnelly.1972, Neidig.Cliver.1983}. Thus, optical continuum is presumably present in all flares but only in a few cases it reaches a measurable degree of brightness. This conclusion implies that WLFs are not fundamentally different from conventional flares \citet{Neidig1989}. Nonetheless, WLFs are important in flare studies because they are similar to stellar flares ways \citet{worden1983} and because they represent the most extreme conditions encountered in solar optical flares \citet{Neidig1989}. Using observations from the Transiting Exoplanet Survey Satellite (TESS), \citet{Ilin2021} present four fully convective stars that exhibited white light flares of large size and long duration. The underlying flare amplitude as a fraction of the quiescent flux of two flares is greater than two. After the discovery of the largest amplitude flares ever recorded on the L0 dwarf, which reached $\Delta V \approx -11$  magnitude \citet{Schmidt2016}, \citet{Jackman2019} detected a large amplitude white-light super-flare on the L2.5 dwarf ULAS J224940.13011236.9 with the flux $\Delta V \approx -10$ magnitude which corresponds to a relative brightness ratio of 10000. This can be demonstrated as follows:
\begin{equation}
	\Delta{V}= V_{\rm max}-V_{\rm min}\approx-10,
\end{equation}
where $V_{\rm max}$ is the apparent magnitude in the visible band corresponding to the flux at the maximum amplitude ($F_{\rm max}$), and $V_{\rm min}$ is the apparent magnitude in the visible band corresponding to the flux at the minimum state ($F_{\rm min}$). Using the magnitude difference calculation :
\begin{equation}
	V_{\rm max}-V_{\rm min} = -10 =2.5 \log_{10} (F_{\rm min}/F_{\rm max}),
\end{equation}
it is clear that $F_{\rm max}/F_{\rm min} = {\Delta{F}}/{F} = 10000$.

Stellar superflares have been studied in multiple wavelength bands, such as X-rays and the H$\alpha$ band. It is important to mention relevant studies here:  \citet{Wu2022} analyzed spectroscopic data from LAMOST DR7 and identified a stellar flare on an M4-type star that is characterized by an impulsive increase followed by a gradual decrease in the H$\alpha$ line intensity. The H$\alpha$ line, which corresponds to a specific transition in hydrogen, exhibits a Voigt profile during the flare. After the impulsive increase in the H$\alpha$ line intensity, a clear enhancement was observed in the red wing of the H$\alpha$ line profile. Additionally, the estimated total energy radiated through the H$\alpha$ line during the flare is on the order of $10^{33}$ erg, providing an indication of the overall energy release associated with the event. Chandra/HETGS time-resolved X-ray spectroscopic observations were used by \citet{Chen2022} to study the behaviour of stellar flares on EV Lac. They discovered distinct plasma flows caused by flares in the corona of EV Lac, but none of them provided evidence for the actual occurrence of stellar CMEs. In most flares, the flow of plasma is accompanied by a rise in the density and temperature of the coronal plasma.

Here we present the detection of 14 super-flares on 13 slowly-rotating Sun-like stars, 12 and 7 cases of large amplitude super-flares on five G-type dwarfs and four M-type dwarfs respectively. Section \ref{section:Method2} presents the method used including the flare detection, the flare energy estimation, and rotation period determination. Section \ref{section:results2} provides the main results of this study. Section \ref{section:conclusion2} closes this work by providing our main conclusions.

\section{Methods}\label{section:Method2}
\subsection{Flare Detection}
\begin{figure*}
	\centering
	\begin{subfigure}{\linewidth}
		\mbox{{\includegraphics[width=0.5\textwidth]{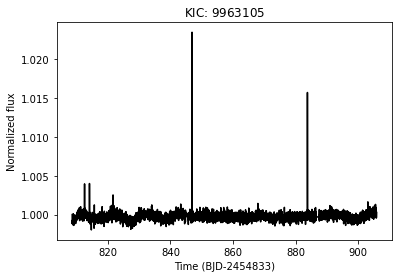}}\quad
		{\includegraphics[width=0.5\textwidth]{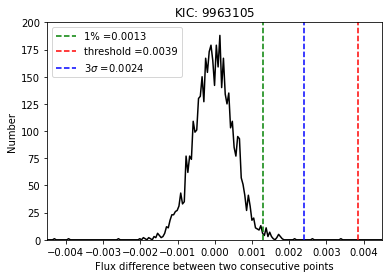} }}
	\end{subfigure}
	\caption{Illustrations of flares detection method used by \citet{Shibayama_2013}. (a) The light curve of KIC 9963105. (b) The distribution of brightness variation between each pair of adjacent data points in the light curves of KIC 9963105. The green vertical line represents the value of 1\% of the total area under the curve, the red vertical lines represents the flare detection threshold and the blue vertical represents $ 3\sigma $ of the brightness variation distribution.} \label{threshold}	
\end{figure*}

We conducted an automated search for super-flares on main-sequence stars type (A, F, G, K, M) based on entire Kepler data, using our Python script on long cadence data from Data Release 25 (DR 25) following \citep{maehara2012,Shibayama_2013} method. The parameters for all targets observed by Kepler have been taken from the Kepler Stellar interactive table in NASA Exoplanet Archive. The study was carried out on a sample of main sequence stars, comprising 2222, 10307, 25442, 10898, and 2653 stars for the spectral types of M, K, G, F, and A, respectively. The following is a brief description of this method. We generate light curves of the stars using the PDCSAP flux. Then, in order to be statistically precise, we computed the distributions of brightness variation by calculating the flux difference in adjacent time intervals between every two neighboring data points in the light curve. Then, we determine the flux difference value at which the area under the distribution equals 1\% of the total area. In order to increase the threshold, the 1\% value of the area was multiplied by a factor of three. The start time of a flare was defined as the time at which the flux difference between two consecutive points exceeded the threshold for the first time. To determine the end time of the flare, we computed the three standard deviations 3$\sigma$ of the distribution of brightness variation. Figure \ref{threshold} displays a typical results for this method for KIC 9963105. The light curve of KIC 9963105 are shown in Figure \ref{threshold}(a). Figure \ref{threshold}(b) shows the distribution of the brightness difference between every two adjacent data points of KIC 9963105 light curve. 1\% of the total area under the distribution curve is represented by the green vertical line. The red vertical line represents the flare detection threshold value, which is equal to three times 1\% value of the area under the distribution curve. The blue vertical line is 3$\sigma $ of the distribution of the brightness variation. To determine the flare end time, we fit a B-spline curve through three points on the relative flux ($\Delta F / F_{\rm  avg}$) distributed around the flare. One point just before the flare, and the other two points five and eight hours after the flare peak, respectively. Then we subtract the B-spline curve from the relative flux in order to remove long-term brightness variations around the flare \citet{Shibayama_2013}. We define the flare end time as the time when the relative flux produced by the subtraction drops below the value of 3$\sigma $ for the first time. After detecting flare events, conditions were applied to all flare candidates. These conditions are: the flare duration must exceed 0.05 days, corresponding to at least three data points two of them after the flare peak, and the flare's decline phase must be longer than its rising phase \citet{Shibayama_2013}. Only flare incidents meeting these criteria were analysed.

\subsection{Energy Calculation}
\citet{Schaefer00} identified nine cases of super-fares involving $10^{33}-10^{38}$ ergs on normal solar-type stars. Their super-flare energy estimate has a large uncertainty, e.g. Groombridge 1830 (HR 4550) total flare energy (in the blue band alone) is $10^{35}$ ergs with an uncertainty of a factor of a few due to having only four points on the light curve. 

The possibility that super-flares can be explained by magnetic energy stored on the star's surface was considered by \citet{Nogami14} Using the Ca ii 8542 absorption line, they estimate that the average magnetic field strength ($B$) of KIC 9766237 and KIC 9944137 is 1-20 Gauss, and the super-flare of these targets has a total energy of $10^{34}$ erg. Under the assumption that the energy released during the flare represents a fraction ($f$) of the magnetic energy stored around the spot area. Their flare energy ($E_ {\rm flare}$) was calculated as follows:
\begin{equation}
	E_ {\rm flare} \sim f\dfrac{B^{2}}{8\pi}L^{3}.
\end{equation}
The length of the magnetic structure causing the flare ($L$), has been considered to be the same size as the spotted region, i.e. $L =\sqrt{a \pi R_{\ast}^{2}}$ where $a$ is the spot's area, which giving that
\begin{equation}
	E_ {\rm flare} \sim f\dfrac{B^{2}}{8\pi}(a \pi R_{\ast}^{2})^{3/2}.
\end{equation}

Our energy estimation for each flare depends on the star's luminosity ($L_{\rm star}$), flare amplitude ($f_{\rm amp}$), and flare duration \citep{Shibayama_2013, yang2017}.
$L_{\rm star}$, the amount of energy that the star emits in one second, is proportional to the star's radius $R$ squared and its surface temperature $T_{\rm eff}$ to the fourth power, and is obtained from the following equation:
\begin{equation}
	L_{\rm star} = \sigma_{\rm  SB}T_{\rm  eff}^4 4\pi R^2, 
\end{equation}
where $ \sigma_{\rm  SB}$ is the Stefan-Boltzmann constant, $4\pi R^2$ is the entire surface area of the star. The continuum emission from a white-light flare is consistent with blackbody radiation at around 9000 K, as suggested by \citep{Hawley1992,Kretzschmar2011}. Based on \citep{Shibayama_2013, yang2017, gunther2020}, we set $T_{\rm  flare}$ = 9000 K and derive the luminosity of a blackbody-emitting star as follows:
\begin{equation} \label{Lflare}
	L_{\rm  flare}(t) = \sigma_{\rm  SB}T^4_{\rm  flare}A_{\rm  flare},
\end{equation}
where $A_{\rm  flare}$ is the flare's area, as determined by the formula:
\begin{equation} \label{Aflare}
	A_{\rm  flare}(t) = f_{\rm  amp}(t) \pi R^2 \frac{\int R_{\lambda}B_{\lambda }(T_{\rm  eff}) d\lambda}
	{\int R_{\lambda}B_{\lambda }(T_{\rm  flare}) d\lambda},
\end{equation}
where $f_{\rm  amp}$ represents the flare amplitude for the relative flux and $R_{\lambda}$ represents the Kepler instrument's response function \citet{caldwell2010kepler}. The Kepler photometer covers various wavelengths, from 420 to 900 nm. The Plank function at a specific wavelength, denoted by $B_{\lambda}(T)$, is given by:
\begin{equation}
	B_{\lambda}(T) = \frac{{2hc^2}/{\lambda^5}}{e^{{hc}/{\lambda k T}}-1},	
\end{equation} 
where $h$ represents Planck's constant, $c$ the speed of light, $T$ the black body temperature, and $k$ Boltzmann's constant.
By substituting Eq.(\ref{Aflare}) into (\ref{Lflare}), we calculate the total flare energy by the integral of $L_{\rm  flare}$ over the flare duration :
\begin{equation} \label{Eflare}
	E_{\rm  flare} =\int_{t_{\rm start}}^{t_{\rm end}} L_{\rm  flare}(t)dt.
\end{equation}

We determine energy of the flares using \citet{Shibayama_2013} energy estimation method, which assumes blackbody radiation from both the star and flare, with a fixed flare temperature of 10,000 K, to estimate the quiescent luminosity. 
We note that \citet{Shibayama_2013} energy estimation
 can have an error of up to 60\% and yet this is more accurate  
than the one used by \citet{Schaefer00} and \citet{Nogami14}.
To improve the accuracy,
\citet{Davenport_2016} proposed an alternative method for estimating the quiescent luminosity of each star to determine the actual energy of the flares. They used the Equivalent Duration (ED) parameter, which represents the integral under the flare in fractional flux units, as a relative energy measurement for each flare event without requiring flux calibration of the Kepler light curves. To calculate the actual energy of the flares emitted in the Kepler band pass (erg), the ED values (sec) are multiplied by the quiescent luminosity (erg/sec) of the respective star. This approach establishes an absolute scale for the relative flare energies, as the quiescent luminosity is individually estimated for each star.

\subsection{Rotational Period Determination}
Light curve periods were calculated using the Lomb-Scargle periodogram, a common statistical approach for detecting and characterising periodic signals in sparsely sampled data. We used an oversampling factor of five \citet{VanderPlas2018}, and use PDCSAP flux to generate a Lomb-Scargle periodogram for each light curve from Q2 to Q16. Furthermore, the period corresponding to the maximum power of the periodogram was allocated as the rotation period for the Kepler ID in a specific quarter. This value was estimated with an accuracy of a day without the decimal component because fractions of a day would not significantly alter the results, allowing us to automate the selection of the star's rotation period rather than manually. We set 0.5 days for periods shorter than a day and eliminated periods less than 0.1 days. Finally, for each Kepler ID, we choose the most frequent period across all quarters from Q2 to Q16. Following the \citet{McQuillan2014} technique, we required that the period chosen for all quarters be identified in at least two unique segments, with the segment defined as three consecutive Kepler quarters. (Q2,Q3,Q4), (Q5, Q6, Q7), (Q8, Q9, Q10), (Q11, Q12, Q13) and (Q14,Q15,Q16).

\section{Results}\label{section:results2}
By performing an automated search for super-flares on G-type main-sequence stars during 1442 days of Kepler observation in all of (DR 25) long-cadence data from Q0 to Q17, we found 14 super-flares on 13 slowly rotating Sun-like stars in each of (KIC 3124010, KIC 3968932, KIC 7459381, KIC 7459381, KIC 7821531, KIC 9142489, KIC 9528212, KIC 9963105, KIC 10275962, KIC 11086906, KIC 11199277, KIC 11350663 and KIC 11971032), with a surface temperature of $5600 K \leqslant T_{\rm  eff} < 6000 K$, a surface gravity of $\log \ g > 4.0$, and a rotational period $P_{\rm rot}$ range between 24.5 and 44 days. Figure \ref{Sun-like} shows seven light curves of these events. The left panels display light curves over a 90-day of observation. The blue arrow on the left panel indicates the observed super-flare, which met all conditions. The right panels show zoomed in time light curves of super-flares. The blue squares represent the data points for a super-flare. We fitted an exponential decay function to the flare light curve to characterise the flares, shown by a red dashed curve. This exponential decay function is given by:
\begin{equation}
	f(t) = a \ e^{-t/\tau} + b
\end{equation}
where $f(t)$ is the relative flux as a function of time, $a$ is the flare peak
height, which is approximately equal to the relative flux at the flare peak, $b$ is the relative flux in the quiescent state and $\tau$ is the decay time of the flare, which is the time at which the relative flux is decreased to $1/e \simeq$ 0.3679 of its initial value. The value of $\tau$, $a$ and $b$ of the exponential decay function for each flare are shown in the right panel.
\begin{figure}[!htbp]
	\centering
	\begin{subfigure}{\linewidth}
		\mbox{{\includegraphics[width=0.5\textwidth]{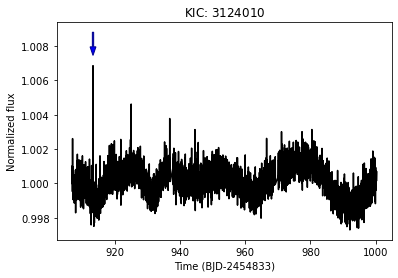}}\quad
			{\includegraphics[width=0.5\textwidth]{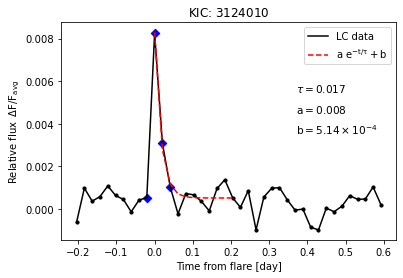} }}
		\mbox{{\includegraphics[width=0.5\textwidth]{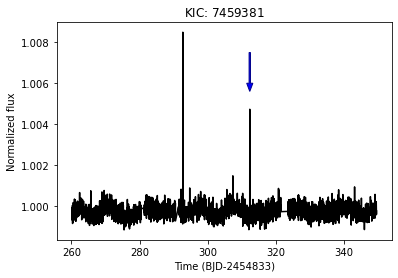}}\quad
			{\includegraphics[width=0.5\textwidth]{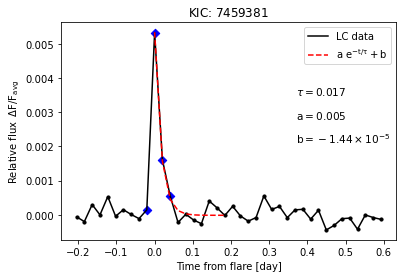} }}
		\mbox{{\includegraphics[width=0.5\textwidth]{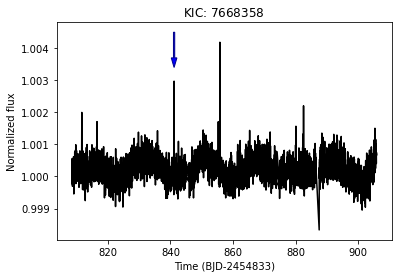}}\quad
			{\includegraphics[width=0.5\textwidth]{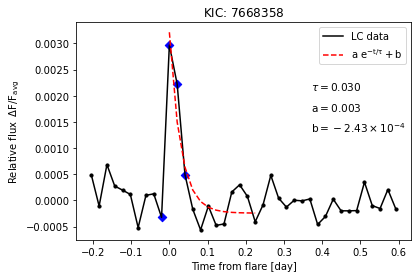} }}
		\mbox{{\includegraphics[width=0.5\textwidth]{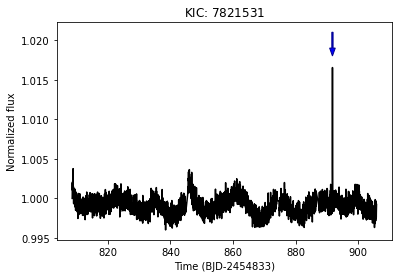}}\quad
			{\includegraphics[width=0.5\textwidth]{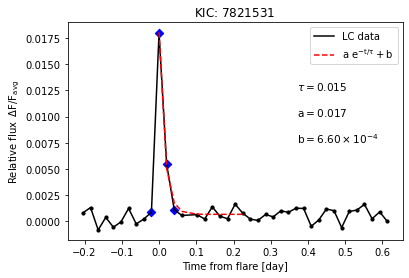} }}
	\end{subfigure}
	\caption{The left panel displays the light curves of super-flares. The x-axis in the left panel is the time in (BJD) and the y-axis is the normalized flux. The blue arrows indicates the occurrence of
		super-flares. The right panels show zoom in time of these super-flares. The x-axis in the right panel is the time from the flare peak in (day) and the y-axis is the relative flux ($\rm \Delta {F}/F_{avg}$). Each data points for a super-flare is represented by blue squares in the right panel. The dashed red curve indicates an exponential fit of the decay phase. $\tau$ in the equation refers to the best fit of exponential decay time, $a$ refers to the final value of the amplitude fit and $b$ refers to the fitted relative flux in the quiescent state.} \label{Sun-like}		
\end{figure}
\begin{figure}[!htbp]\ContinuedFloat
	\centering
	\begin{subfigure}{\linewidth}
		\mbox{{\includegraphics[width=0.5\textwidth]{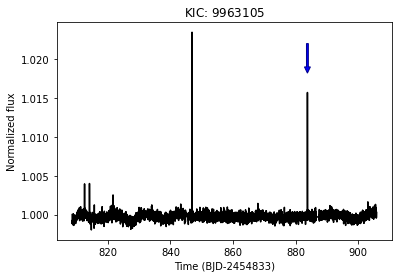}}\quad
			{\includegraphics[width=0.5\textwidth]{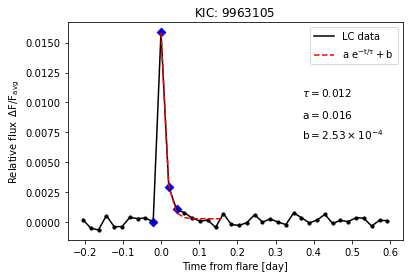} }}
		\mbox{{\includegraphics[width=0.5\textwidth]{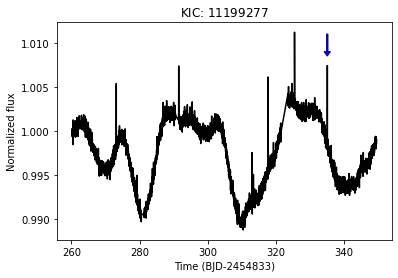}}\quad
			{\includegraphics[width=0.5\textwidth]{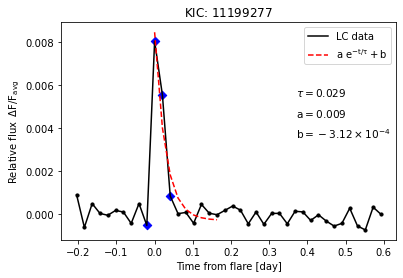} }}
		\mbox{{\includegraphics[width=0.5\textwidth]{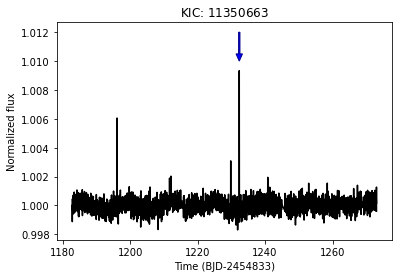}}\quad
			{\includegraphics[width=0.5\textwidth]{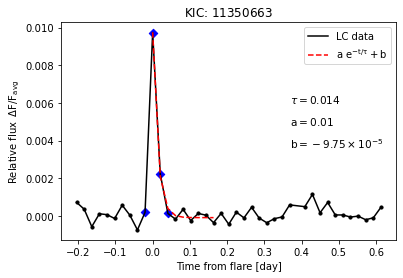} }}
	\end{subfigure}
	\caption{Figure \ref{Sun-like} continued} \label{Sun-like}	
\end{figure}
Flare parameters and their duration, amplitudes, energies and $\tau$ values are listed in Table \ref{tab:slowly rotating}. The rotation periods for these slowly rotating Sun-like stars were taken from \citet{McQuillan2014}. Their flare energies range from $(1.9-9.0)\times 10^{34}$ erg. The flare amplitude of the slowly rotating Sun-like stars is relatively small, ranging between 0.002 and 0.018. We also find that the duration of flares in all of these cases is the same 0.061 days. This is probably due to the fact that flare duration of all small amplitude super-flares is two data points in time or \textit{less}. Therefore, because one of the flare detection conditions in our code is that there must be at least two data points between the flare's peak and the end, cases with flare duration less than two points were not detected, and we end up with the same duration super-flares with the two data points. Because it appears that there are no small amplitude super-flares with duration \textit{greater} than two data points, all flare duration end up the same. This selection effect is because we use long cadence light curve data, which has a 29.4-minute interval between each data point in time. The value of $\tau$ varies from 0.012 to 0.036 days.
\begin{table*}[!htbp]
	\centering
	\scriptsize
	\caption{super-flares on slowly rotating Sun-like stars.}
	\label{tab:slowly rotating}
	\begin{adjustbox}{width=1\textwidth}
	\begin{tabular*}{\linewidth}{ @{\extracolsep{\fill}}cccccccccccc}
		\hline
		\hline
		\text{Kepler ID} & $\rm T_{eff}$& \text{log g} & \text{Radius} & $P_{\rm rot}$$^{\rm a}$ & $\rm t_{start}$ & $\rm t_{end}$ & $\rm t_{peak}$ &\text{amp} & \text{Flare Duration}& $\rm \tau$ & \text{Flare Energy} \\
		& ($K$) & &($R_\odot$) & \text{(day)}& \text{(BJD)} &\text{(BJD)}&\text{(BJD)}&&\text{(day)} &\text{(day)}& \text{(erg)}\\
		\hline
		3124010 & 5688 & 4.46 &1.01& 25.90 & 913.24 & 913.30 & 913.26 & 0.008 & 0.061 & 0.017 & $4.57\times 10^{34}$\\
		3968932 & 5716 & 4.39 & 0.96 & 24.56 & 868.88 & 868.94 & 868.90 & 0.004 & 0.061 & 0.026 & $3.89 \times 10^{34}$\\
		7459381 & 5635 & 4.27 & 1.11 & 26.19 & 312.31 & 312.37 & 312.33 & 0.005 & 0.061 & 0.017 &$4.89 \times  10^{34}$\\
		7668358 & 5668 & 4.36 & 0.98 & 41.83 & 841.13 & 841.19 & 841.15 & 0.003 & 0.061 & 0.030 & $ 1.94 \times 10^{34}$\\
		7821531 & 5681 & 4.52 & 0.92 & 32.66 & 891.81 & 891.87 &891.83 & 0.018 & 0.061 & 0.015 & $8.63 \times 10^{34}$\\
		9142489 & 5878	& 4.51	& 0.95	& 25.20	& 1561.13 & 1561.19 & 1561.15 & 0.004 & 0.061 & 0.028 & $ 2.61 \times 10^{34}$\\
		9528212 & 5872 & 4.42 & 0.97 & 61.43 & 1332.34 & 1332.40 & 1332.36 & 0.003 & 0.061 & 0.036 & $ 2.31 \times 10^{34}$\\
		9963105 & 5751 & 4.39 & 1.01 & 28.09 & 883.80 & 883.86 & 883.82 & 0.016 & 0.061 & 0.012 & $ 9.00 \times 10^{34}$\\
		10275962 & 5782 & 4.51 & 0.91 & 26.12 & 213.21 & 213.27 & 213.23 & 0.007 & 0.061 & 0.035 & $ 3.71 \times 10^{34}$\\
		10275962 & 5782 & 4.51 & 0.91 & 26.12 & 599.85 & 599.91 & 599.87 & 0.004 & 0.061 & 0.029 & $ 2.04 \times 10^{34}$\\
		11086906 & 5758 & 4.38 & 1.11 & 29.18 & 1206.01 & 1206.07 & 1206.03 & 0.002 & 0.061 & 0.018 & $ 1.90 \times 10^{34}$\\
		11199277 & 5638 & 4.49 & 0.92 & 29.00 & 325.43 & 325.49 & 325.45 & 0.008 & 0.061 & 0.029 & $ 3.72 \times 10^{34}$\\
		11350663 & 5966 & 4.49 & 0.96 & 36.92 & 1232.31 & 1232.37 & 1232.33 & 0.010 & 0.061 & 0.014 & $ 5.21 \times 10^{34}$\\
		11971032 & 5942 & 4.51 & 0.94 & 44.00 & 1231.92 & 1231.98 & 1231.94 & 0.006 & 0.061 & 0.030 & $ 3.80 \times 10^{34}$\\
		\hline
	\end{tabular*}
	\end{adjustbox}
	\begin{tablenotes}
		\small
		\item 
		\ $^{\rm a}$ Rotation period from \citet{McQuillan2014}.
	\end{tablenotes}	
\end{table*}

\begin{figure}[!htbp]
	\centering
	\includegraphics[width=0.6\textwidth]{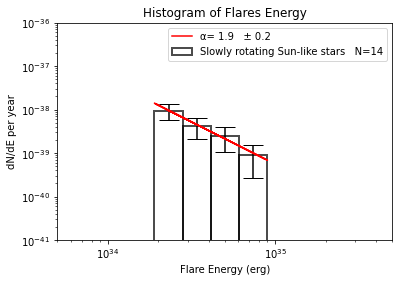}
	\caption{log-log scale histograms showing the distribution of flare frequency as a function of flare energy of 14 super-flares on slowly rotating Sun-like stars. The distribution follows a power-law relation $dN/dE \propto E^{-\alpha}$ where $\alpha = 1.9\pm 0.2$} 
	\label{occurrence of superflares}	
\end{figure}

We calculated the frequency distribution of the 14 super-flares on the 13 slowly rotating Sun-like stars and plotted a log scale histogram presenting this distribution as shown in Figure \ref{occurrence of superflares}. The x-axis represents the flare's energy, and the y-axis represents the number of super-flares per star per year per unit of energy. Therefore, we calculated the weight for each bin using
\begin{equation}
	w =\frac{3.16\times10^{7}}{N_{\rm  os}\times D\times E}, 
\end{equation}
where $N_{\rm os}$ is the number of observed stars, $D$ is the duration of the observation period in seconds, and $E$ is the super-flare energy that belongs to that bin. From the number of stars in Table 3 in our previous work \citet{paper1}, we estimated that the number of observed G-type dwarfs with $5600 K \leqslant T_{\rm eff} < 6000 K$ and $P_{\rm rot} > 10$ days is equal to 19160 stars. Since this distribution is related to slowly rotating Sun-like stars with $P_{\rm rot}$ between 24.5 and 44 days, we estimated the number of the observed stars to be one-third of the original sample, i.e. 5635 stars, given that the average rotation period is 34 days which is almost three times the period of 10 days. We estimated the probability of the occurrence of super-flares in slowly rotating Sun-like stars with $P_{\rm rot}$ of 24.5 to 44 days. We found that the rate of super-flares incidence with the energy of $4.54 \times 10^{34}$ erg is $1.94 \times 10^{-4}$ flares per year per star, corresponding to a super-flare occurring on a star once every 5160 years. We calculated this value by multiplying the average energy from the x-axis by the average dN/dE from the y-axis, $4.54 \times 10^{34} \times 4.27\times 10^{-39} = 1.94 \times 10^{-4} $ flares per year per star, and by taking the reciprocal of $1.94 \times 10^{-4}$, we get 5160 which gives the number of years in which a flare occurs on a star. The frequency distribution of these 14 super-flares follows a power law relation $dN/dE \propto E^{-\alpha}$ where $\alpha = 1.9\pm 0.2$. This is consistent with our previous result in \citet{paper1} for the frequency distribution of slowly rotating G-type dwarfs where $\alpha = 2.0\pm 0.1$.

In addition to the 14 cases of of super-flares on slowly rotating Sun-like starts, we detected 12 super-flares
with a large amplitude on 6 G-type dwarfs in each of KIC 5865248, KIC 6783223, KIC 7505473, KIC 10053146, KIC 10057002 and KIC 11709752. Figure \ref{large-amp} shows eight light curves of these events same as Figure \ref{Sun-like}. Table \ref{tab:larg-amp} shows the duration, amplitude, energy, and $\tau$ values for these super-flares with their parameters. The energy of their flares range from $1.67\times 10^{36}$ to $1.42 \times 10^{38}$ erg. The rotation period of KIC 1170952 was obtained by this work. As for the rotation period for the other five stars, no such data is available. Even applying the method described in Paper I does not allow period determination in these five cases. According to \citet{yang2017}, there are three possible reasons: (i) due to the inclination angle and low activity level, the light curve has a small amplitude at the accuracy level of Kepler; (ii) fast-rotating stars have spots in the poles \citet{schussler1992}, making it hard to detect light variation through rotation; and (iii) the rotation period is longer than 90 days (a quarter), making it difficult (or impossible) to detect them in the frequency spectrum of the star. The flare amplitude for these cases range between 4.05 and 35.60. These flares tend to last longer than flares with smaller amplitude of slowly rotating Sun-like stars as their duration varies between 0.061 and 0.143 day. The $\tau$ values of flares exhibiting large amplitude on G-type main-sequence stars are observed to be higher than those of flares occurring on slowly rotating Sun-like stars, as their values range between 0.014 and 0.058 days.
\begin{figure}[!htbp]
	\centering
	\begin{subfigure}{\linewidth}
		\mbox{{\includegraphics[width=0.5\textwidth]{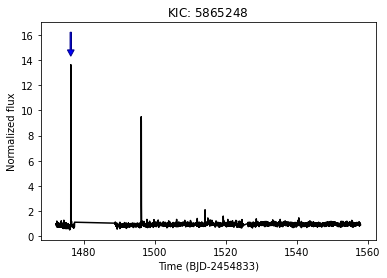}}\quad
			{\includegraphics[width=0.5\textwidth]{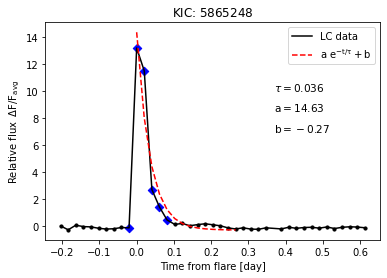} }}
		\mbox{{\includegraphics[width=0.5\textwidth]{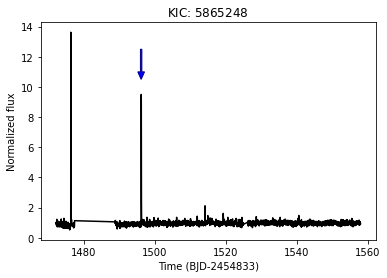}}\quad
			{\includegraphics[width=0.5\textwidth]{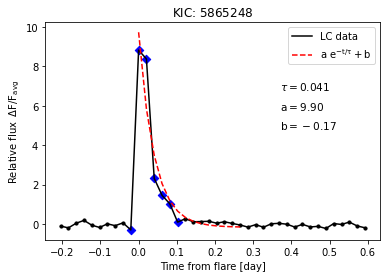} }}
		\mbox{{\includegraphics[width=0.5\textwidth]{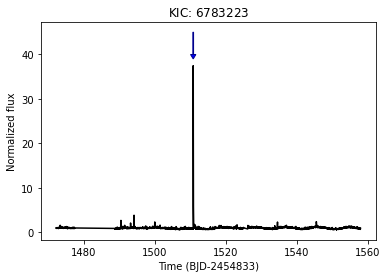}}\quad
			{\includegraphics[width=0.5\textwidth]{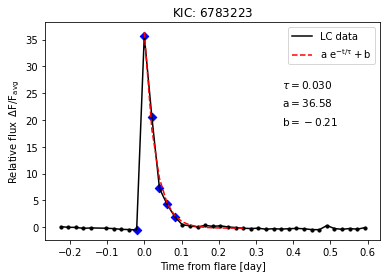} }}
		\mbox{{\includegraphics[width=0.5\textwidth]{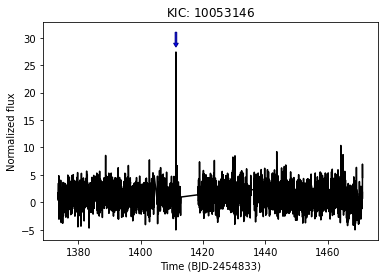}}\quad
			{\includegraphics[width=0.5\textwidth]{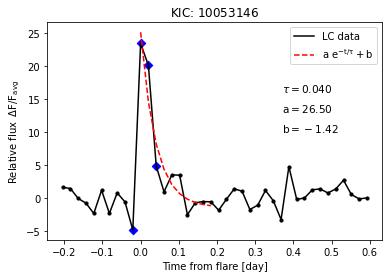} }}
	\end{subfigure}
	\caption{Same as Figure \ref{Sun-like} but for large amplitude super-flares on G-type main-sequence stars.}	
	\label{large-amp}	
\end{figure}

\begin{figure}[!htbp]\ContinuedFloat
	\centering
	\begin{subfigure}{\linewidth}
		\mbox{{\includegraphics[width=0.5\textwidth]{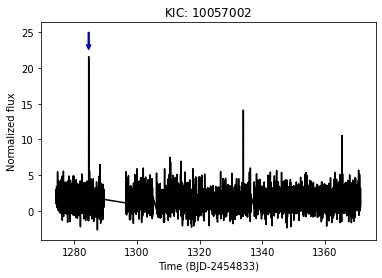}}\quad
			{\includegraphics[width=0.5\textwidth]{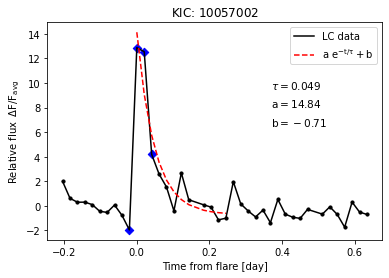} }}
		\mbox{{\includegraphics[width=0.5\textwidth]{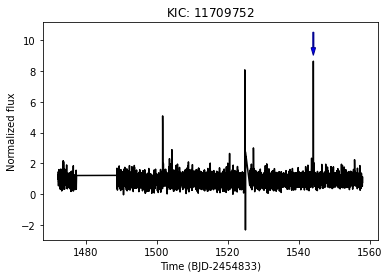}}\quad
			{\includegraphics[width=0.5\textwidth]{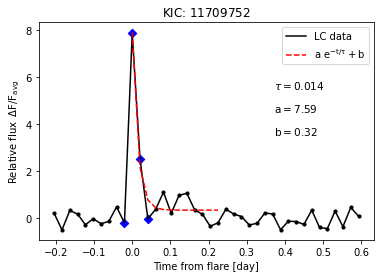} }}
		\mbox{{\includegraphics[width=0.5\textwidth]{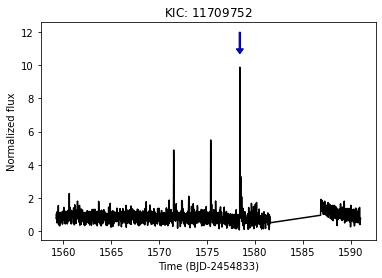}}\quad
			{\includegraphics[width=0.5\textwidth]{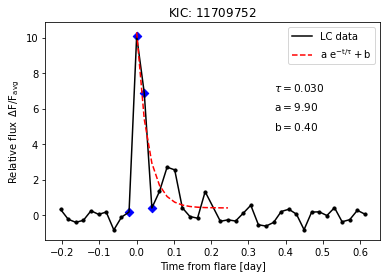} }}
	\end{subfigure}
	\caption{Same as Figure \ref{Sun-like} but for large amplitude super-flares on G-type main-sequence stars} 
	\label{large-amp}	
\end{figure}

\begin{table*}[!htbp]
	\centering
	\scriptsize
	\caption{Large amplitude super-flares on G-type main-sequence stars.}
	\label{tab:larg-amp}
	\begin{adjustbox}{width=1\textwidth}
	\begin{tabular*}{\linewidth}{ @{\extracolsep{\fill}}cccccccccccc}
		\hline
		\hline
		\text{Kepler ID} & $\rm T_{eff}$ & \text{log g} & \text{Radius} & $P_{\rm rot}$$^{\rm a}$ & $\rm t_{start}$ & $\rm t_{end}$ & $\rm t_{peak}$ &\text{amp} & \text{Flare Duration}& $\rm \tau$ & \text{Flare Energy} \\
		& ($K$) & &($R_\odot$) & \text{(day)}& \text{(BJD)} &\text{(BJD)}&\text{(BJD)}&&\text{(day)} &\text{(day)}& \text{(erg)}\\
		\hline
		5865248&5780&4.44&1&NA&1476.22&1476.33&1476.24&13.16&0.102&0.036&$7.29\times 10^{37}$\\
		5865248&5780&4.44&1&NA&1496.09&1496.21&1496.11&8.78&0.123&0.041&$9.56\times 10^{37}$\\
		5865248&5780&4.44&1&NA&1561.78&1561.86&1561.80&8.14&0.082&0.034&$6.75\times 10^{37}$\\
		6783223&5780&4.44&1&NA&1510.76&1510.86&1510.78&35.60&0.102&0.030&$8.6\times 10^{37}$\\
		7505473&5780&4.44&1&NA&1385.81&1385.95&1385.85&4.05&0.143&0.058&$1.67\times 10^{36}$\\
		10053146&5780&4.44&1&NA&1411.27&1411.33&1411.29&22.75&0.061&0.040&$1.42\times 10^{38}$\\
		10057002&5780&4.44&1&NA&1284.52&1284.58&1284.54&12.03&0.061&0.049&$6.24\times 10^{37}$\\
		11709752&5780&4.44&1&0.5&1501.56&1501.62&1501.58&4.21&0.061&0.031&$2.85\times 10^{37}$\\
		11709752&5780&4.44&1&0.5&1544.06&1544.13&1544.08&7.65&0.061&0.014&$2.85\times 10^{37}$\\
		11709752&5780&4.44&1&0.5&1571.49&1571.55&1571.51&4.48&0.061&0.033&$3.94\times 10^{37}$\\
		11709752&5780&4.44&1&0.5&1575.35&1575.43&1575.37&5.11&0.082&0.046&$5.44\times 10^{37}$\\
		11709752&5780&4.44&1&0.5&1578.39&1578.45&1578.41&9.40&0.061&0.030&$3.94\times 10^{37}$\\		
		\hline
	\end{tabular*}
	\end{adjustbox}
	\begin{tablenotes}
		\small
		\item 
		\ $^{\rm a}$ Rotation period from \citet{paper1}.
	\end{tablenotes}
\end{table*}

For stars of other spectral classes, no significant flares with large amplitudes were detected on main-sequence stars of type A, F, and K. Only M-type main-sequence stars manifested seven super-flares with large amplitude on each of KIC 6580019, KIC 7123391, KIC 7341517 and KIC 9201463. Similar to Figures \ref{Sun-like} and \ref{large-amp}, Figure \ref{M-type-large-amp} displays the seven light curves for these events. The parameters of these super-flares, including their duration, amplitude, energy, and $\tau$ values, are displayed in Table \ref{tab:M-type-lamp}. These flares have an energy between $3.16 \times 10^{33}$ and $1.59 \times 10^{35}$ erg, amplitude ranges between 3.91 and 15.14 and their duration lasts between 0.018 and 0.044 day. $\tau$ values for super-flares with large amplitude on M-type main sequence stars vary from 0.030 to 0.049 days.

\begin{table*}[!htbp]
	\centering
	\scriptsize
	\caption{Large amplitude super-flares on M-type main-sequence stars.}
	\label{tab:M-type-lamp}
	\begin{adjustbox}{width=1\textwidth}
	\begin{tabular*}{\linewidth}{ @{\extracolsep{\fill}}cccccccccccc}
		\hline
		\hline
		\text{Kepler ID} & $\rm T_{eff}$ & \text{log g} & \text{Radius} & $P_{\rm rot}$ & $\rm t_{start}$ & $\rm t_{end}$ & $\rm t_{peak}$ &\text{amp} & \text{Flare Duration}& $\rm \tau$ & \text{Flare Energy} \\
		& ($K$) & &($R_\odot$) & \text{(day)}& \text{(BJD)} &\text{(BJD)}&\text{(BJD)}&&\text{(day)} &\text{(day)}& \text{(erg)}\\
		\hline
		6580019 & 2661 & 5.28 & 0.12 & NA & 609.97 & 610.09 & 609.99 & 3.91 & 0.123 & 0.044 & 7.04$\times  10^{33}$\\
		6580019 & 2661 & 5.28 & 0.12 & NA & 674.45 & 674.60 & 674.47 & 10.30 & 0.143 & 0.041 & 1.25$\times  10^{34}$\\
		7123391 & 3326 & 5.12 & 0.19 & NA & 638.53 & 638.59 & 638.55 & 8.07 & 0.061 & 0.038 & 4.76$\times  10^{34}$\\
		7123391 & 3326 & 5.12 & 0.19 & NA & 692.09 & 692.19 & 692.13 & 12.38 & 0.102 & 0.018 & 1.23$\times  10^{35}$\\
		7123391 & 3326 & 5.12 & 0.19 & NA & 794.54 & 794.72 & 794.56 & 15.14 & 0.184 & 0.049 & 1.59$\times  10^{35}$\\
		7341517 & 2661 & 5.28 & 0.12 & NA & 877.95 & 878.12 & 877.99 & 5.27 & 0.163 & 0.041 & 4.82$\times  10^{33}$\\
		9201463 & 3319 & 5.14 & 0.18 & NA & 215.11 & 215.27 & 215.15 & 5.51 & 0.163 & 0.030 & 3.16$\times  10^{33}$\\
		\hline
	\end{tabular*}
	\end{adjustbox}
\end{table*}

\begin{figure}[!htbp]
	\centering
	\begin{subfigure}{\linewidth}
		\mbox{{\includegraphics[width=0.5\textwidth]{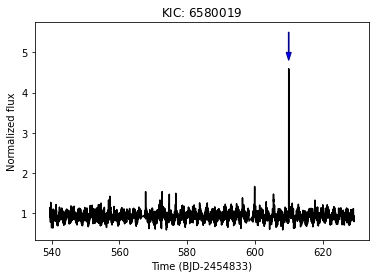}}\quad
			{\includegraphics[width=0.5\textwidth]{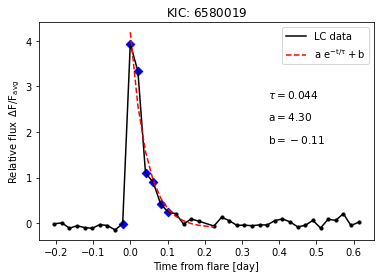} }}
		\mbox{{\includegraphics[width=0.5\textwidth]{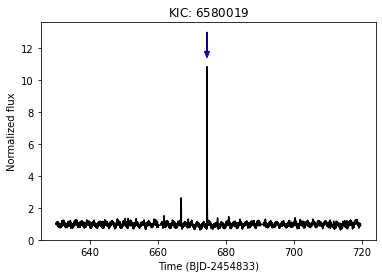}}\quad
			{\includegraphics[width=0.5\textwidth]{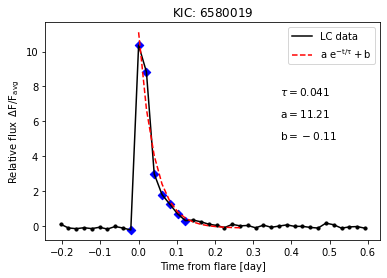} }}
		\mbox{{\includegraphics[width=0.5\textwidth]{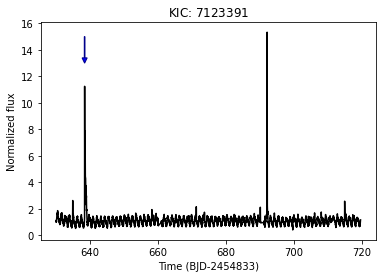}}\quad
			{\includegraphics[width=0.5\textwidth]{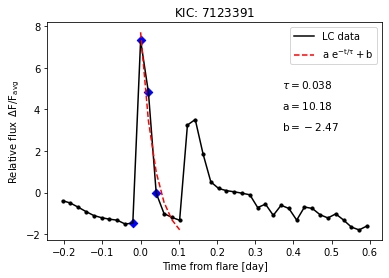} }}
		\mbox{{\includegraphics[width=0.5\textwidth]{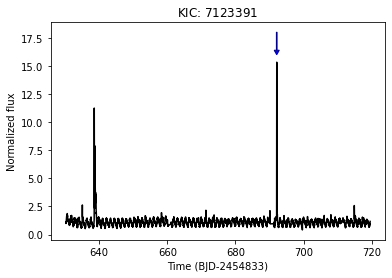}}\quad
			{\includegraphics[width=0.5\textwidth]{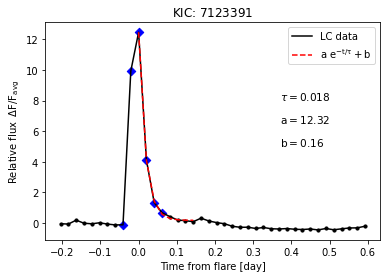} }}
	\end{subfigure}
	\caption{Same as Figures \ref{Sun-like} and \ref{large-amp} but for large amplitude super-flares on M-type main-sequence stars.} \label{M-type-large-amp}		
\end{figure}

\begin{figure}[!htbp]\ContinuedFloat
	\centering
	\begin{subfigure}{\linewidth}
		\mbox{{\includegraphics[width=0.5\textwidth]{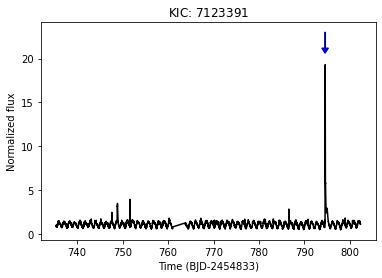}}\quad
			{\includegraphics[width=0.5\textwidth]{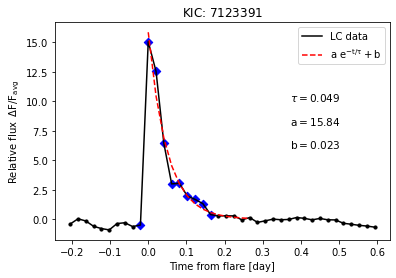} }}
		\mbox{{\includegraphics[width=0.5\textwidth]{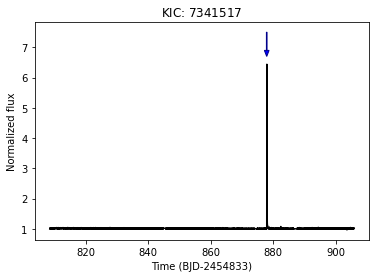}}\quad
			{\includegraphics[width=0.5\textwidth]{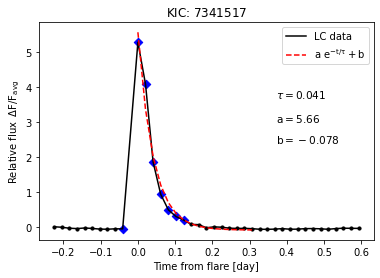} }}
		\mbox{{\includegraphics[width=0.5\textwidth]{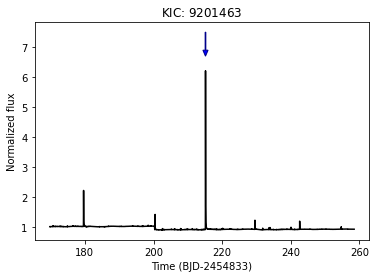}}\quad
			{\includegraphics[width=0.5\textwidth]{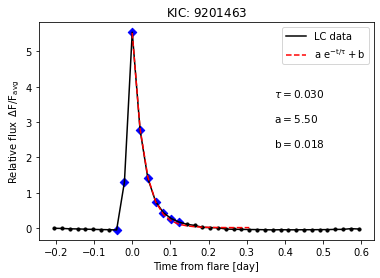} }}
	\end{subfigure}
	\caption{Same as Figures \ref{Sun-like} and \ref{large-amp} but for large amplitude super-flares on M-type main-sequence stars.} \label{M-type-large-amp}	
\end{figure}
\begin{figure}[!htbp]
	\centering
	\begin{subfigure}{\linewidth}
		\mbox{{\includegraphics[width=0.5\textwidth]{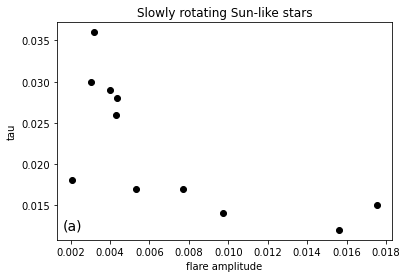}}\quad
			{\includegraphics[width=0.5\textwidth]{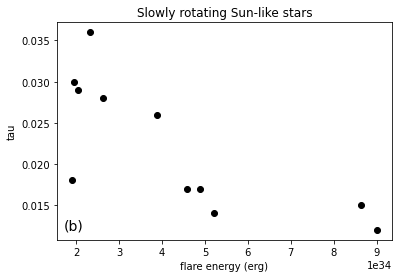} }}
		\mbox{{\includegraphics[width=0.5\textwidth]{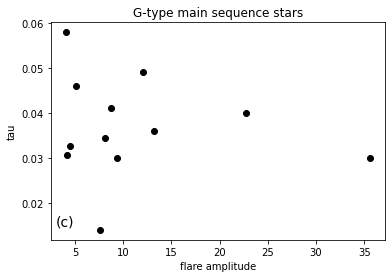}}\quad
			{\includegraphics[width=0.5\textwidth]{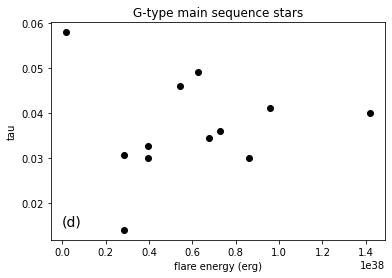} }}
		\mbox{{\includegraphics[width=0.5\textwidth]{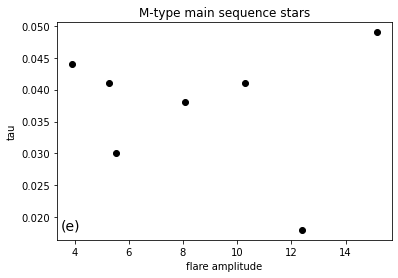}}\quad
			{\includegraphics[width=0.5\textwidth]{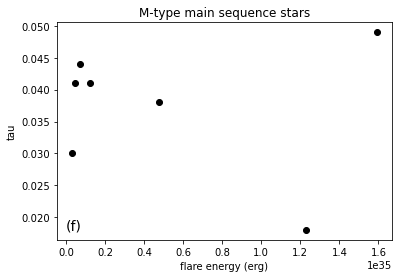} }}
	\end{subfigure}
	\caption{The left panels display a scatter plot showing the relation between tau values on the y-axis with the flare amplitude $f_{\rm amp}$ on the x-axis. While the right panels display a scatter plot showing the relation between $\tau$ values on the y-axis with the flare energy $E_{\rm flare}$ on the x-axis. For flares on slowly rotating Sun-like stars,(a) demonstrates that $\tau$ values are large for low flare amplitudes but consistently small for high flare amplitudes. Likewise,(b) demonstrates that high $\tau$ values correspond to low flare energies, whereas low $\tau$ values correspond to high flare energies. For large amplitude super-flares on G-type dwarfs (c,d) and M-type dwarfs (e,f), $\tau$ has no clear connection to the $f_{\rm amp}$ or $E_{\rm flare}$.  } \label{tau-amp-energy}	
\end{figure}

We examined whether there is a dependence between $\tau$ vs. flare amplitude ($f_{\rm amp}$) and $\tau$ vs. flare energy ($E_{\rm flare}$). Therefore, we graphically display six panels in Figure \ref{tau-amp-energy} showing the relationship between $\tau$ and the amplitude of flares and $\tau$ and the energy of flares in slow-rotating Sun-like stars \ref{tau-amp-energy}(a, b), G-type stars \ref{tau-amp-energy}(c, d) and M-type stars \ref{tau-amp-energy}(e, f) respectively. In \ref{tau-amp-energy}(a) for slowly-rotating Sun-like stars, we find that for small amplitude, $\tau$ is large, and when the amplitude is large, $\tau$ is consistently small in the
range considered. The same applies to the relation between $\tau$ and energy in Figure \ref{tau-amp-energy}(b), we see that large $\tau$ values correspond to small energies and small values of $\tau$ correspond to large energies considered. On the contrary, there is no clear relation between $\tau$ vs. $f_{\rm amp}$ and $\tau$ vs. $E_{\rm flare}$ in G-type and M-type main sequence stars in Figure \ref{tau-amp-energy}(c-f). However, 
as mentioned in the Introduction, 
according to \citet{Maehara2015}, the duration of 
superflares, $\tau$, scales as the flare energy, 
$E$, according to $\tau \propto E^{0.39\pm 0.03}$. 
Similarly, \citet{Tu2020} found that 
 $T_{duration}\propto E^{0.42  \pm 0.01}$.
It broadly follows from the simple reconnection scaling arguments, 
that $\tau \propto E^{1/3}$. We believe that we could not deduce such scaling because of small number of data points in Figure \ref{tau-amp-energy}. We tried various functions of fit using Python's \textit{curve\_fit} and Excel's \textit{trendline}, referred to as a (line of best fit), to visualize the general trend for the data. We could not find any reliable, functional fit dependence between those parameters because \textit{the coefficient of determination}, $\rm R^{2}$, which shows how well the data fit the regression model, is less than 0.5 for all those cases in Figure \ref{tau-amp-energy}(a to f). Hence any attemted fit has been unreliable, as only fit with $\rm R^{2} > 0.5$ can be deemed acceptable. To determine the extent to which the two variables, $\tau$ and flare amplitude, as well as $\tau$ and flare energy, are correlated, we calculated the \textit{Pearson Correlation Coefficient} ($r$), which measures the strength and direction of the relationship between two variables, using
IDL's built-in function $\rm CORRELATE(X,Y)$ and Python's function \textit{scipy.stats.pearsonr}. Both IDL and Python gave the same values.
The values for the Pearson correlation coefficient ($r$)  between two datasets are listed in Table \ref{tab:Pearson}. We note that 
for the slowly rotating sun-like stars datasets, 
 $r= -0.592$ and $r= -0.691$ for $\tau$ vs. $f_{amp}$ 
 and $\tau$ vs. $E_{flare}$ respectively, which suggests a 
 noticable negative correlation between the variables.
For the remaining 4 cases
these $r$ values are close to zero, 
which indicates a weak or nonexistent 
correlation between the two variables.
In general, $r$ varies from $-1$ to $1$.
The extreme cases of $r=\pm 1$ mean that 
there is clear linear correlation/anti-correlation.
$r=0$ means that there is no linear relation between the variables.

\begin{table}[!h]
	\centering
	\caption{The Pearson correlation coefficient between $\tau$ vs. $f_{\rm amp}$ and $\tau$ vs. $E_{\rm flare}$.}
	\label{tab:Pearson}
	\begin{tabular}{ccc}
		\hline
		\hline
		{\rm Sample} & $\tau$ {vs.} $f_{\rm amp}$ & $\tau$ {vs.} $E_{\rm flare}$\\
		\hline
		\rm Slowly rotating sun like stars & -0.592 & -0.691 \\
		\rm G-type large amplitude flares  & -0.149 & 0.141\\
		\rm M-type large amplitude flares & -0.046 & -0.098\\		
		\hline
	\end{tabular}	
\end{table}

In the context of explaining the absence of large-amplitude flares detected in A-, F-, and K-type main-sequence stars, while they are only detected in G-type and M-type stars we would like to remark the following.
Using Kepler space telescope, \citet{Chang2018} studied of the light curves of the M dwarfs.
They found a number of flare events with the peak flux increases
$\Delta F / F \ge 1$.
Magnetic fields of the M dwarfs
are generated by turbulent magnetic dynamo mechanism.
This is due to their deep convective zones, and this
leads to very powerful flares, compared the G-type stars
\citep{Davenport2014}. As for G-type stars, detection of strength of
flares in such stars have been know for some time
starting from \citep{maehara2012}.
Therefore is not entirely surprising that that we detected
large-amplitude flares in G- and M-type stars.
As for A-, F-, K-type stars we remark that according to
\citet{Pedersen2017} for the flare generation, stars must have:
eaither a deep outer convection zone for F5-type and 
perhaps later-types; or
strong, radiatively driven winds for B5-type and
earlier types;
or strong large-scale magnetic
fields for A and B-type stars. 
\citet{Pedersen2017} and earlier works suggest that normal
A-type stars have non such features 
and thus should not flare.
However, flares and super-flares have previously been detected on such
stars according to \citep{Bai2020} and
references therein. The situation with K-type stars is somewhat a 'gray
area'. Stars less massive and cooler than our Sun are K dwarfs; and even
fainter and cooler stars are the red-coloured M dwarfs. 
Thus K dwarfs are
probably borderline case where 
large-amplitude flares can occur.

\section{Conclusions}\label{section:conclusion2}
Using our Python script on long cadence data from Data Release 25 (DR 25), we searched for super-flares on main-sequence stars of types (A, F, G, K, and M) based on the entire Kepler data following the method of \citep{maehara2012,Shibayama_2013}. The Kepler targets' parameters were retrieved from the Kepler Stellar interactive table in the NASA Exoplanet Archive. Using these data, we detected 14 super-flare on 13 Sun-like stars with a surface temperature of $5600 K \leqslant T_{\rm  eff} < 6000 K$, and $P_{\rm rot}$ range from 24.5 to 44 days. In addition, we found 12 and 7 cases of large amplitude super-flares on six and four main-sequence G and M type stars, respectively. Main-sequence stars of other spectral types A, F, and K showed no signs of large-amplitude super-flares. To characterise the flares, we fit an exponential decay function to the flare light curve given by $f(t) = a \ e^{-t/\tau} + b$. We study the relation between the decay time of the flare after
its peak $\tau$ vs. $f_{\rm amp}$ and  $\tau$ vs. $E_{\rm flare}$. For slowly rotating Sun-like stars, we find that $\tau$ is large for small flare amplitudes and $\tau$ is small for large flare amplitudes considered. Similarly, we find that large $\tau$ values correspond to small flare energies and small $\tau$ values correspond to high flare energies considered. However, for the main sequence stars of the G and M types, $\tau$ has no apparent relation to the $f_{\rm amp}$ or $E_{\rm flare}$. We experimented with several different fit functions between $\tau$ vs. $f_{\rm amp}$ and  $\tau$ vs. $E_{\rm flare}$ to better see the underlying pattern in the data. Since the $\rm R^{2}$ is less than 0.5 in these cases, we could not identify a reliable fit functional dependence between these parameters.\\

In conclusion, we believe that:\\ 
(i) the thirteen peculiar Kepler IDs that are Sun-like, slowly rotating with rotation periods of 24.5 to 44 days, and yet can produce a super-flare with
energies in the range of $(2\textendash9)\times 10^{34}$ erg; and\\
(ii) six G-type and four M-type Kepler IDs with exceptionally large amplitude super-flares, with the relative flux in the range 
$\Delta F / F_{\rm  avg} = 4-35$; \\
defy our current understanding of stars and hence are worthy of further investigation.

\section*{Acknowledgements}
Some of the data presented in this paper were obtained from the Mikulski Archive for Space Telescopes (MAST). STScI is operated by the Association of Universities for Research in Astronomy, Inc., under NASA contract NAS5-26555. Support for MAST for non-HST data is provided by the NASA Office of Space Science via grant NNX13AC07G and by other grants and contracts.\\
Authors would like to thank Deborah Kenny of STScI for kind assistance in obtaining the data, Cozmin Timis and Alex Owen of Queen Mary University of London for the assistance in data handling at the Astronomy Unit.\\
A. K. Althukair wishes to thank Princess Nourah Bint Abdulrahman University, Riyadh, Saudi Arabia and  
Royal Embassy of Saudi Arabia Cultural Bureau in London, UK for the financial support of her PhD scholarship, held at Queen Mary University of London.

Authors would like to thank an anonymous referee whose comments
greatly improved this manuscript.

\section*{Data Availability}
All data used in this study was generated by our bespoke Python script that can be found at \url{https://github.com/akthukair/AFD} under the filename AFD.py and other files in the same GitHub repository.
The data underlying this article were accessed from Mikulski Archive for Space Telescopes (MAST) \url{https://mast.stsci.edu/portal/Mashup/Clients/Mast/Portal.html}. The long-cadence Kepler light curves analyzed in this paper can be accessed via MAST \citet{https://doi.org/10.17909/t9488n}. The Kepler Stellar parameters table for all targets can be found at \citet{https://doi.org/10.26133/nea6}. The derived data generated in this research will be shared on reasonable request to the corresponding author.

\bibliography{msRAA-2023-0185.R1}{}
\bibliographystyle{aasjournal}

\end{document}